\documentclass[usletter,11pt]{article}
\usepackage{amsfonts}
\usepackage{amsmath}
\usepackage{amssymb}
\usepackage{amsthm}
\usepackage{mathrsfs}
\usepackage{marvosym}

\newcommand{\g}{g}

\DeclareMathOperator*{\Exp}{\mathbf{E}}
\DeclareMathOperator*{\Prob}{\mathbf{P}}
\DeclareMathOperator{\adeg}{\deg_{1/3}}
\DeclareMathOperator{\degeps}{\deg_{\epsilon}}

\DeclareMathOperator{\disc}{disc}

\newcommand{\ma}{\text{\it MA}}
\newcommand{\NP}{\mathsf{NP}}
\newcommand{\coNP}{\mathsf{coNP}}
\newcommand{\MA}{\mathsf{MA}}
\newcommand{\AC}{\mathsf{AC}}
\newcommand{\BPP}{\mathsf{BPP}}
\newcommand{\e}{\mathrm{e}}   %
\newcommand{\moon}{\{-1,+1\}^n}
\newcommand{\zook}{\zoo^k}
\newcommand{\zoom}{\zoo^m}
\newcommand{\zoon}{\{0,1\}^n}
\newcommand{\moo}{\{-1,+1\}}
\newcommand{\zoo}{\{0,1\}}
\newcommand{\Z}{\mathbb{Z}}
\renewcommand{\geq}{\geqslant}
\renewcommand{\leq}{\leqslant}
\renewcommand{\Re}{\mathbb{R}}
\newcommand{\boldchi}{\underline{\chi\!}}
\newcommand{\boldsigma}{\mbox{\underbar{$\sigma\!$}}}
\long\def\symbolfootnote[#1]#2{\begingroup%
\def\thefootnote{\fnsymbol{footnote}}\footnote[#1]{#2}\endgroup} 
\newenvironment{restatetheorem}[1]{\begin{trivlist}\item[\hskip
\labelsep{\bf Theorem~\ref{#1}}  {(Restated from p.~\pageref{#1}).}]\it}
{\end{trivlist}}

\theoremstyle{plain}
\newtheorem{theorem}{Theorem}[section]

\newtheorem{proposition}[theorem]{Proposition}
\newtheorem{corollary}[theorem]{Corollary}
\newtheorem{claim}[theorem]{Claim}
\newtheorem{fact}[theorem]{Fact}

\theoremstyle{definition}

\theoremstyle{remark}

\newtheorem*{remark*}{Remark}

\numberwithin{equation}{section}

\begin{document}

\sloppy

\title{A Separation of $\NP$ and $\coNP$ in Multiparty Communication
Complexity\vspace{5mm}}

\date{}

\author{
{\sc Dmitry Gavinsky}\thanks{
NEC Laboratories America Inc.,
4 Independence Way, Suite 200,
Princeton, NJ 08540.
}
\and
{\sc Alexander~A.~Sherstov}\thanks{Microsoft Research,
Cambridge, MA 02142.
Email: $\mathtt{sherstov@cs.utexas.edu}$
}
}

\setcounter{page}{0}
\maketitle
\thispagestyle{empty}

\begin{abstract}
We prove that $\NP\ne\coNP$ and $\coNP\nsubseteq\MA$ in the
number-on-forehead model of multiparty communication complexity for
up to $k=(1-\epsilon)\log n$ players, where $\epsilon>0$ is any
constant.  Specifically, we construct a function $F:(\zoon)^k\to\zoo$
with co-nondeterministic complexity $O(\log n)$ and Merlin-Arthur
complexity $n^{\Omega(1)}.$ The problem was open for $k\geq3.$
\end{abstract}

\section{Introduction} \label{sec:intro}

The number-on-forehead model of multiparty communication
complexity~\cite{cfl83multiparty} features $k$ communicating players
whose goal is to compute a given distributed function. More precisely,
one considers a Boolean function $F:(\zoon)^k\to\moo$ whose arguments
$x_1,\dots,x_k\in\zoon$ are placed on the foreheads of players~$1$
through $k,$ respectively. Thus, player~$i$ sees all the arguments
except for $x_i.$ The players communicate by writing bits on a
shared blackboard, visible to all. Their goal is to compute
$F(x_1,\dots,x_k)$ with minimum communication.
The multiparty model
has found a variety of applications, including circuit complexity,
pseudorandomness, and proof
complexity~\cite{yao90multiparty,hastad-goldman91multiparty,bns92,razborov-wigderson93multiparty,bps07lovasz-schrijver}.
This model draws its richness from the 
overlap in the players' inputs, which makes it challenging to prove
lower bounds. 
Several fundamental questions in the multiparty model remain open despite much research.

\subsection{Previous Work and Our Results}
The $k$-party number-on-forehead model naturally gives rise to the
complexity classes $\NP_k^{cc},$ $\coNP_k^{cc},$ $\BPP_k^{cc},$ and
$\MA_k^{cc}$, corresponding to communication problems $F:(\zoon)^k\to\moo$
with efficient nondeterministic, co-nondeterministic, randomized,
and Merlin-Arthur protocols, respectively.  An efficient protocol
is one with communication cost $\log^{O(1)} n.$ Determining the
exact relationships among these classes is a natural goal in
complexity theory.

For example, 
it had been open to show that nondeterministic protocols can
be more powerful than randomized, for $k\geq3$ players.
This problem was recently solved
in~\cite{lee-shraibman08disjointness,chatt-ada08disjointness} for
up to $k=(1-o(1))\log_2\log_2 n$ players, and later strengthened 
in~\cite{pitassi08np-rp} to $k=(1-\epsilon)\log_2n$ players,
where $\epsilon>0$ is any given constant. An
explicit separation for the latter case was obtained 
in~\cite{david-pitassi-viola08bpp-np}.

The contribution in this paper is to 
relate the power of nondeterministic, co-nondeterministic, and
Merlin-Arthur protocols.  For $k=2$ players, the relations among
these models are well understood~\cite{ccbook,klauck03rect}:
it is known that $\coNP_2^{cc}\ne\NP_2^{cc}$ and
further that $\coNP_2^{cc}\nsubseteq\MA_2^{cc}.$
Starting at $k=3,$ however, it has been open to even separate
$\NP_k^{cc}$ and $\coNP_k^{cc}$.
Our main result is that $\coNP_k^{cc}\nsubseteq\MA_k^{cc}$ for up
to $k=(1-\epsilon)\log_2n$ players, where $\epsilon>0$ is an arbitrary
constant.  The separation is by an explicitly given function.  In
particular, our work shows that $\NP_k^{cc}\neq\coNP_k^{cc}$ and
also subsumes the separation in~\cite{pitassi08np-rp,
david-pitassi-viola08bpp-np}, since $\NP_k^{cc}\subseteq\MA_k^{cc}$
and $\BPP_k^{cc}\subseteq\MA_k^{cc}$.  Let the symbols $N(F)$,
$N(-F)$, and $\ma(F)$ denote the nondeterministic, co-nondeterministic,
and Merlin-Arthur complexity of $F$ in the $k$-party number-on-forehead
model.

\newcommand{\mainthe}
{Let $k\leq (1-\epsilon)\log_2n,$ where $\epsilon>0$ is any given constant.  
Then there is an {\rm (}explicitly given{\rm )} function $F:(\zoon)^k\to\moo$ with 
\begin{align*}
N(-F)=O(\log n)
\end{align*}
and
\begin{align*}
\ma(F)= n^{\Omega(1)}.
\end{align*}
In particular, 
$\coNP_k^{cc}\nsubseteq\MA_k^{cc}$ and $\NP_k^{cc}\ne\coNP_k^{cc}.$}

\noindent
\parbox{\textwidth}{
\begin{theorem}[Main Result]
\mainthe
\label{thm:main}
\end{theorem}
}

\bigskip

It is a longstanding open problem to exhibit a
function with nontrivial multiparty complexity for $k\geq\log_2n$ 
players.  Therefore, the separation in Theorem~\ref{thm:main} is state-of-the-art with
respect to the number of players.

The proof of Theorem~\ref{thm:main}, to be described
shortly, is based on the \emph{pattern matrix
method}~\cite{sherstov07ac-majmaj,sherstov07quantum} and its multiparty
generalization in~\cite{david-pitassi-viola08bpp-np}.
In the final section
of this paper, we revisit several other multiparty
generalizations~\cite{arkadev07multiparty, 
lee-shraibman08disjointness,
chatt-ada08disjointness,
beame-huyn-ngoc08multiparty-circuits}
of the pattern matrix method.
By applying our techniques in these other settings, we are able to obtain similar exponential separations by functions as simple as constant-depth circuits. 
However, these new separations only hold up to
$k=\epsilon\log n$ players, unlike the separation in Theorem~\ref{thm:main}.

\subsection{Previous Techniques}

Perhaps the best-known method for communication lower bounds, both
in the number-on-forehead multiparty model and various two-party
models, is the \emph{discrepancy method}~\cite{ccbook}. 
The method consists in exhibiting a distribution $P$ with
respect to which the function $F$ of interest has negligible
discrepancy, i.e., negligible correlation with all low-cost protocols.
A more powerful technique is the \emph{generalized
discrepancy method}~\cite{klauck01quantum,razborov03quantum}. This
method consists in exhibiting a distribution $P$ and a function $H$
such that, on the one hand, the function $F$ of interest is
well-correlated with $H$ with respect to $P,$ but on the other
hand, $H$ has negligible discrepancy with respect to $P.$

In practice, considerable effort is required to find
suitable $P$ and $H$ and to analyze the resulting
discrepancies.  In particular, no strong bounds were
available on the discrepancy or generalized discrepancy of constant-depth
circuits $\AC^0.$  The recent \emph{pattern matrix
method}~\cite{sherstov07ac-majmaj,sherstov07quantum}
solves this problem for $\AC^0$ and a large family of
other matrices. More specifically, the method
uses standard analytic properties of Boolean functions
(such as approximate degree or threshold degree) to
determine the discrepancy and generalized discrepancy of
the associated communication problems.

Originally formulated in
\cite{sherstov07ac-majmaj,sherstov07quantum} for the two-party model,
the pattern matrix method  has been adapted to the multiparty model by 
several authors~\cite{arkadev07multiparty,
lee-shraibman08disjointness,
chatt-ada08disjointness,
pitassi08np-rp,
david-pitassi-viola08bpp-np,
beame-huyn-ngoc08multiparty-circuits}.
The first adaptation
of the method to the multiparty model gave improved lower
bounds for the multiparty disjointness
function~\cite{lee-shraibman08disjointness,chatt-ada08disjointness}.
This line of work was combined in~\cite{pitassi08np-rp,
david-pitassi-viola08bpp-np}
with probabilistic arguments to separate
the classes $\NP_k^{cc}$ and $\BPP_k^{cc}$ for up to
$k=(1-\epsilon)\log_2n$ players, by an explicit
function.  A new 
paper~\cite{beame-huyn-ngoc08multiparty-circuits} 
gives polynomial lower bounds for constant-depth
circuits, in the model with up
to $k=\epsilon\log n$ players. Further details on this body of
research and other duality-based
approaches~\cite{shi-zhu07block-composed} can be found in the survey
article~\cite{dual-survey}.

\subsection{Our Approach}

To obtain our main result, we combine the work in~\cite{pitassi08np-rp,
david-pitassi-viola08bpp-np} with several new ideas. First, we
derive a new criterion for high nondeterministic
communication complexity, inspired by the Klauck-Razborov generalized
discrepancy method~\cite{klauck01quantum, razborov03quantum}. 
Similar to Klauck-Razborov, we also look for a
hard function $H$ that is well-correlated with the function $F$ of
interest, but we additionally quantify the agreement of $H$ and $F$
on the set $F^{-1}(-1).$ This agreement ensures that
$F^{-1}(-1)$ does not have a small cover by cylinder intersections,
thus placing $F$ outside $\NP_k^{cc}.$ To handle 
the more powerful Merlin-Arthur model, we combine
this development with an earlier technique~\cite{klauck03rect}
for proving lower bounds against two-party Merlin-Arthur protocols.

In keeping with the philosophy of the pattern matrix method, we
then reformulate the agreement requirement for $H$ and $F$ as a
suitable analytic property of the underlying Boolean function $f$ and
prove this property directly, using linear programming duality. 
The function $f$ in question happens to be OR.

Finally, we apply our program to the specific function $F$ constructed
in~\cite{david-pitassi-viola08bpp-np} for the purpose of separating
$\NP_k^{cc}$ and $\BPP_k^{cc}.$ Since $F$ has small nondeterministic
complexity by design, the proof of our main result is complete once
we apply our machinery to $-F$ and derive a lower bound on
$\ma(-F).$

\subsection{Organization}
We start in Section~\ref{sec:prelim} with relevant
technical preliminaries and standard background on multiparty
communication complexity.  In Section~\ref{sec:discrepancy}, we
review the original discrepancy method, the generalized discrepancy
method, and the pattern matrix method.  In Section~\ref{sec:newmethod},
we derive the new criterion for high nondeterministic
and Merlin-Arthur communication complexity.  The proof
of Theorem~\ref{thm:main} 
comes next, in Section~\ref{sec:main}.  In the final section
of the paper, we explore some implications of this work in light of
other multiparty papers~\cite{arkadev07multiparty, lee-shraibman08disjointness,
chatt-ada08disjointness, beame-huyn-ngoc08multiparty-circuits}.

\section{Preliminaries} \label{sec:prelim}

We view Boolean functions as mappings $X\to\moo,$ where $X$ is a
finite set such as $X=\zoon$ or $X=\zoon\times\zoon.$ We identify $-1$ and
$+1$ with ``true'' and ``false,'' respectively. The notation
$[n]$ stands for the set $\{1,2,\dots,n\}.$ For integers $N,n$ with
$N\geq n,$ the symbol ${[N]\choose n}$ denotes the family of all
size-$n$ subsets of $\{1,2,\dots,N\}.$ For a string $x\in\moo^N$
and a set $S\in{[N]\choose n},$ we define
$x|_S=(x_{i_1},x_{i_2},\dots,x_{i_n})\in\moon,$ where $i_1<i_2<\cdots<i_n$
are the elements of $S.$ For $x\in\zoon,$ we write $|x| =
x_1+\cdots+x_n.$ 
Throughout this manuscript, ``$\log$'' refers to
the logarithm to base~$2.$ For a function $f:X\to\Re,$ where $X$
is an arbitrary finite set, we write $\|f\|_\infty=\max_{x\in X}
|f(x)|.$

We will need the following observation regarding discrete
probability distributions on the hypercube, cf.~\cite{sherstov07ac-majmaj}.
\begin{proposition}
Let $\mu(x)$ be a probability distribution on $\zoon.$ Fix
$i_1,\dots,i_n$ $\in\{1,2,\dots,n\}.$ Then
\[\sum_{x\in\zoon} \mu(x_{i_1},\dots,x_{i_n})\leq
2^{n-|\{i_1,\dots,i_n\}|}.\]
\label{prop:distribution}
\end{proposition}

For functions
$f,\g:X_1\times\cdots\times X_k\to\Re$ (where $X_i$ is a finite set,
$i=1,2,\dots,k$), we define $\langle f,\g\rangle =
\sum_{(x_1,\dots,x_k)} f(x_1,\dots,x_k)\g(x_1,\dots,x_k).$ 
When $f$ and $\g$ are vectors or matrices,
this is the standard definition of inner product. The \emph{Hadamard
product} of $f$ and $\g$ is the tensor $f\circ \g:X_1\times\cdots\times
X_k\to\Re$ given by 
$(f\circ \g)(x_1,\dots,x_k)=f(x_1,\dots,x_k)\g(x_1,\dots,x_k).$

The symbol $\Re^{m\times n}$ refers to the family of all $m\times
n$ matrices with real entries.  The $(i,j)$th entry of a matrix $A$
is denoted by $A_{ij}.$ In most matrices that arise in this work,
the exact ordering of the columns (and rows) is irrelevant. In such
cases, we describe a matrix using the notation $[F(i,j)]_{i\in I,\,
j\in J},$ where $I$ and $J$ are some index sets.

We conclude with a review of the Fourier transform over $\Z_2^n.$ 
Consider the vector space of functions $\zoon\to\Re,$ equipped with
the inner product
$\langle f,\g\rangle = 2^{-n} \sum f(x)\g(x).$
For $S\subseteq[n],$ define $\chi_S:\zoon\to\moo$ by
$\chi_S(x) =(-1)^{\sum_{i\in S} x_i}.$
Then $\{\chi_S\}_{S\subseteq[n]}$ is an orthonormal basis for the
inner product space in question.  As a result, every function
$f:\zoon\to\Re$ has a unique representation of the form
$f=\sum_{S\subseteq[n]} \hat f(S)\,\chi_S,$ where $\hat
f(S)=\langle f,\chi_S\rangle$.  The reals $\hat f(S)$ are
called the \emph{Fourier coefficients} of $f.$
The following fact is immediate from the definition of $\hat f(S)$:

\begin{proposition}
Fix $f:\zoon\to\Re.$ Then 
\[\max_{S\subseteq[n]}|\hat f(S)| 
\leq 2^{-n} \sum_{x\in\zoon} |f(x)|.\]
\label{prop:fourier-coeff-bound}
\end{proposition}

\subsection{Communication Complexity}  \label{sec:cc}

An excellent reference on communication complexity is the monograph
by Kushilevitz and Nisan~\cite{ccbook}. In this overview, we will
limit ourselves to key definitions and notation.
The simplest model of communication in this work is the
two-party randomized model.  Consider a function $F:X\times Y\to\moo,$
where $X$ and $Y$ are finite sets. Alice receives an input
$x\in X,$ Bob receives $y\in Y,$ and their objective is to predict
$F(x,y)$ with high accuracy. To this end, Alice and Bob share a
communication channel and have an unlimited supply of shared random
bits.  Alice and Bob's protocol is said to have error $\epsilon$
if on every input $(x,y)$, the computed output differs from the
correct answer $F(x,y)$ with probability no greater than $\epsilon.$
The \emph{cost} of a given protocol is the maximum number of bits
exchanged on any input. The \emph{randomized} communication complexity
of $F,$ denoted $R_\epsilon(F),$ is the least cost of an $\epsilon$-error
protocol for $F.$ It is standard practice to use the shorthand
$R(F)=R_{1/3}(F).$ Recall that the error probability of a protocol
can be decreased from $1/3$ to any other positive constant at the
expense of increasing the communication cost by a constant factor.
We will use this fact in our proofs without further
mention. 

A generalization of two-party communication is 
the \emph{multiparty number-on-forehead} model of communication.
Here one considers a function
$F:X_1\times\cdots\times X_k\to\moo$ for some finite sets $X_1,\dots,X_k.$
There are $k$ players. A given input $(x_1,\dots,x_k)\in
X_1\times\cdots\times X_k$ is distributed among the players by
placing $x_i$ on the forehead of player $i$ (for $i=1,\dots,k$).
In other words, player $i$ knows
$x_1,\dots,x_{i-1},x_{i+1},\dots,x_k$ but not $x_i.$
The players communicate by writing bits on a shared blackboard,
visible to all. They additionally have access to a shared source
of random bits.  Their goal is to devise a communication protocol
that will allow them to accurately predict the value
of $F$ on every input. 
Analogous to the two-party case,
the randomized communication complexity $R_\epsilon(F)$ is 
the least cost of an $\epsilon$-error
communication protocol for $F$ in this model, and $R(F)=R_{1/3}(F)$. 

Another model in this paper is the number-on-forehead
\emph{nondeterministic} model. 
As before, one considers a function
$F:X_1\times\cdots\times X_k\to\moo$ for some finite sets
$X_1,\dots,X_k.$ An input from $X_1\times\cdots\times X_k$ is
distributed among the $k$ players as before.  
At the start of the protocol, $c_1$ unbiased nondeterministic bits
appear on the shared blackboard.  Given the values of those bits,
the players behave deterministically, exchanging an additional $c_2$
bits by writing them on the blackboard.  A nondeterministic protocol
for $F$ must output the correct answer for \emph{at least one}
nondeterministic choice of the $c_1$ bits 
when $F(x_1,\dots,x_k)=-1$ and for \emph{all}
possible choices when $F(x_1,\dots,x_k)=+1$.  The cost of a
nondeterministic protocol is defined as $c_1+c_2$.  The
\emph{nondeterministic} communication complexity of $F$, denoted
$N(F),$ is the least cost of a nondeterministic protocol for $F.$
The \emph{co-nondeterministic} communication complexity of $F$ is
the quantity $N(-F)$.

The number-on-forehead \emph{Merlin-Arthur} model combines the power
of randomized and nondeterministic models.  Similar to the
nondeterministic case, the protocol starts with a nondeterministic
guess of $c_1$ bits, followed by $c_2$ bits of communication.
However, the communication can be randomized, and the requirement is that
the error probability be at most $\epsilon$ for \emph{at least one} nondeterministic choice when $F(x_1,\dots,x_k)=-1$ and for \emph{all} possible nondeterministic choices when $F(x_1,\dots,x_k)=+1$.
The cost of a protocol is defined as $c_1+c_2.$ The
\emph{Merlin-Arthur} communication complexity of $F$,
denoted $\ma_\epsilon(F)$, is the least cost of an $\epsilon$-error
Merlin-Arthur protocol for $F.$ We put $\ma(F)=\ma_{1/3}(F)$. 
Clearly, $\ma(F)\leq\min\{N(F),R(F)\}$ for every $F$.

Analogous to computational complexity, one defines $\BPP_k^{cc},$
$\NP_k^{cc},$ $\coNP_k^{cc}$, and $\MA_k^{cc}$ as the classes of
functions $F:(\zoon)^k\to\moo$ with complexity $\log^{O(1)} n$ in
the randomized, nondeterministic, co-nondeterministic, and Merlin-Arthur
models, respectively.

\section{Generalized Discrepancy and Pattern Matrices} \label{sec:discrepancy}

A common tool for proving communication lower bounds is the
\emph{discrepancy method.} Given a function $F:X\times Y\to\moo$
and a distribution $\mu$ on $X\times Y,$ the \emph{discrepancy of
$F$ with respect to $\mu$} is defined as
\[ 
\disc_\mu(F) = \max_{\substack{S\subseteq X,\\T\subseteq Y}}
\quad \left| \sum_{x\in S} \sum_{y\in T}  \mu(x,y)F(x,y)\right|. \]
This definition generalizes to the multiparty case as follows.
Consider a function $F:X_1\times\cdots\times X_k\to\moo$ and a distribution
$\mu$ on $X_1\times\cdots\times X_k.$ The \emph{discrepancy of $F$ with
respect to $\mu$} is defined as
\begin{align*}
\disc_\mu(F) =
\max_{\chi} \left|
\sum_{\substack{(x_1,\dots,x_k)\\\;\;\in X_1\times\cdots\times X_k}}
\mu(x_1,\dots,x_k)
F(x_1,\dots,x_k)
\chi(x_1,\dots,x_k)
\right|, 
\end{align*}
where the maximum ranges over functions 
$\chi:X_1\times\cdots\times X_k\to\zoo$ of the form
\begin{align}
\chi(x_1,\dots,x_k)=\prod_{i=1}^k
\phi_i(x_1,\dots,x_{i-1},x_{i+1},\dots,x_k)
\label{eqn:cylinder-intersection}
\end{align}
for some 
$\phi_i:X_1\times\cdots
X_{i-1}\times X_{i+1}\times\cdots X_k\to\zoo,$ $i=1,2,\dots,k.$
A function $\chi$ of the form (\ref{eqn:cylinder-intersection}) is
called a \emph{rectangle} for $k=2$ and a \emph{cylinder intersection}
for $k\geq3.$ Note that for $k=2,$ the multiparty definition of
discrepancy agrees with the one given earlier for the two-party
model. We put
\begin{align*}
\disc(F)=\min_\mu\disc_\mu(F).
\end{align*}

Discrepancy is difficult to analyze as defined. Typically, one uses
the following estimate, derived by repeated applications
of the Cauchy-Schwarz inequality.

\begin{theorem}[\cite{bns92,chung-tetali93,raz2000bns-chung-tetali-criterion}]
Fix $F:X_1\times\cdots\times X_k\to\moo$ and a distribution $\mu$ on
$X_1\times\cdots\times X_k.$ Put
$\psi(x_1,\dots,x_k)=F(x_1,\dots,x_k)\mu(x_1,\dots,x_k).$
Then
\[ \left(\frac{\disc_\mu(F)}{|X_1|\cdots|X_k|}\right)^{2^{k-1}} \leq 
\Exp_{\substack{x_1^0\in X_1\\ x_1^1\in X_1}}\cdots 
\Exp_{\substack{x_{k-1}^0\in X_{k-1} \\x_{k-1}^1\in X_{k-1}}}
\left|\Exp_{x_k\in X_k} \prod_{z\in\zoo^{k-1}}
\psi(x_1^{z_1},\dots,x_{k-1}^{z_{k-1}}, x_k) \right|. \]
\label{thm:disc-bound}
\end{theorem}

\noindent
In the case of $k=2$ parties, there are other ways to estimate the
discrepancy, including the spectral norm of a
matrix (e.g., see~\cite{sherstov07quantum}).

For a function $F:X_1\times\cdots\times X_k\to\moo$ and a distribution
$\mu$ over $X_1\times\cdots\times X_k,$ let $D^{\mu}_{\epsilon}(F)$
denote the least cost of a deterministic protocol for $F$ whose
probability of error with respect to $\mu$ is at most $\epsilon.$
This quantity is known as the \emph{$\mu$-distributional complexity}
of $F.$  Since a randomized protocol can be viewed as a probability
distribution over deterministic protocols, we immediately have that
$R_\epsilon(F)\geq\max_\mu D^{\mu}_\epsilon(F).$ We are now
ready to state the discrepancy method.

\begin{theorem}[Discrepancy method; see~\cite{ccbook}]
For every $F:X_1\times\cdots\times X_k\to\moo,$ every distribution $\mu$ on
$X_1\times\cdots\times X_k,$ and\, $0<\gamma\leq1,$
\[ R_{1/2 - \gamma/2}(F) \geq D^{\mu}_{1/2-\gamma/2}(F) \geq
\log\frac{\gamma}{\disc_\mu(F)}. \]
\label{thm:original-discrepancy-method}
\end{theorem}

\noindent
In words, a function with small discrepancy is hard to compute
to any nontrivial advantage over random guessing, let alone compute
it to high accuracy. 

\subsection{Generalized Discrepancy Method} \label{sec:discrepancy-method}

The discrepancy method is particularly strong in that it gives
communication lower bounds not only for bounded-error protocols but
also for protocols with error vanishingly close to $\frac12.$ This
strength of the discrepancy method is at once a weakness. 
For
example, the disjointness function {\sc disj}$(x,y)=\bigvee_{i=1}^n
(x_i\wedge y_i)$ has a randomized protocol with error $\frac12 -
\Omega\left(\frac1n\right)$ and communication $O(\log n).$
As a result, the disjointness function
has high discrepancy, and no strong lower bounds can be obtained
for it via the discrepancy method. 
Yet it is well-known that {\sc
disj} has communication complexity $\Theta(n)$ in the
randomized model~\cite{KS92disj,razborov90disj} and $\Omega(\sqrt
n)$ in the quantum model~\cite{razborov03quantum} and Merlin-Arthur 
model~\cite{klauck03rect}.

The \emph{generalized discrepancy method} is an
extension of the traditional discrepancy method that avoids the
difficulty just cited.  This technique was first applied by
Klauck~\cite{klauck01quantum} and reformulated
in its current form by Razborov~\cite{razborov03quantum}. The
development in~\cite{klauck01quantum,razborov03quantum} takes place
in the quantum model of communication.  However, the same idea 
works in a variety of models, as illustrated 
in~\cite{sherstov07quantum}.  The version of the generalized
discrepancy method for the two-party randomized model is as follows.

\noindent\parbox{\textwidth}{
\begin{theorem}[{\cite[\S2.4]{sherstov07quantum}}]
Fix a function $F:X\times Y\to\moo$ and $0\leq\epsilon<1/2.$ Then for all
functions $H:X\times Y\to\moo$ and all probability distributions
$P$ on $X\times Y,$
\begin{align*}
R_\epsilon(F) \geq \log
\frac{\langle F, H\circ P\rangle - 2\epsilon}{\disc_P(H)}.
\end{align*}
\label{thm:2party-generalized-disc}
\end{theorem}
}

\noindent
The usefulness of Theorem~\ref{thm:2party-generalized-disc} stems from its
applicability to functions that have efficient protocols with error
close to random guessing, such as $\frac12-\Omega\left(\frac
1n\right)$ for the disjointness function.
Note that one recovers Theorem~\ref{thm:original-discrepancy-method},
the ordinary discrepancy method, by setting $H=F$ in
Theorem~\ref{thm:2party-generalized-disc}.

\begin{proof}[Proof of Theorem~\textup{\ref{thm:2party-generalized-disc}
(adapted from~\cite{sherstov07quantum}, pp.~88--89).}]
Put $c=R_\epsilon(F).$ A public-coin protocol with cost $c$ can be thought
of as a probability distribution on deterministic protocols with cost at
most~$c.$ In particular, there are random variables
$\boldchi_1,\boldchi_2,\dots,\boldchi_{2^c}:X\times Y\to\zoo,$
each a rectangle, as well as random variables
$\boldsigma_1,\boldsigma_2,\dots,\boldsigma_{2^c}\in\moo,$
such that 
\[ \left\|F - \Exp\left[\sum\boldsigma_i\boldchi_i\right]
\right\|_\infty \leq 2\epsilon. \]
Therefore, 
\begin{align*}
\left\langle F- \Exp\left[\sum\boldsigma_i\boldchi_i\right], H\circ P
\right\rangle
\leq 2\epsilon.
\end{align*}
On the other hand,
\begin{align*}
\left\langle F- \Exp\left[\sum\boldsigma_i\boldchi_i\right],
H\circ P \right\rangle
\geq \langle F,H\circ P\rangle - 2^{c}\disc_P(H)
\end{align*} 
by the definition of discrepancy.  The theorem follows at once from
the last two inequalities.
\end{proof}

Theorem~\ref{thm:2party-generalized-disc} extends word-for-word to
the multiparty model, as follows:

\begin{theorem}[\cite{lee-shraibman08disjointness,chatt-ada08disjointness}]
Fix a function $F:X\to\moo$ and $\epsilon\in[0,1/2),$ where
$X=X_1\times\cdots\times X_k.$ Then for all
functions $H:X\to\moo$ and all probability distributions
$P$ on $X,$
\begin{align*}
R_\epsilon(F) \geq \log
\frac{\langle F, H\circ P\rangle - 2\epsilon}{\disc_P(H)}.
\end{align*}
\label{thm:multi-generalized-disc}
\end{theorem}

\begin{proof}
Identical to the two-party case
(Theorem~\ref{thm:2party-generalized-disc}), with the word
``rectangles'' replaced by ``cylinder intersections.''
\end{proof}

\subsection{Pattern Matrix Method} \label{sec:pattern-mat}

To apply the generalized discrepancy method to a given Boolean
function $F,$ one needs to identify a Boolean function $H$ which is
well correlated with $F$ under some distribution $P$ but has
low discrepancy with respect to $P.$ 
The pattern matrix method~\cite{sherstov07ac-majmaj, sherstov07quantum}
is a systematic technique for 
finding such $H$ and $F.$ 
To simplify the exposition
of our main results, we will now review this method and 
sketch its proof.

Recall that the \emph{$\epsilon$-approximate degree} of a function
$f:\zoon\to\Re,$ denoted $\degeps(f),$ is the least degree of a
polynomial $p$ with $\|f-p\|_\infty\leq\epsilon.$ A starting point
in the pattern matrix method is the following dual formulation of
the approximate degree.
\begin{fact}
Fix $\epsilon\geq0.$ Let $f:\zoon\to\Re$ be given with $d=
\degeps(f)\geq1.$ Then there is a function $\psi:\zoon\to\Re$ such that:
\begin{align*}
&\;\,\hat\psi(S)=0 & \text{for $|S|<d,$}\\
&\sum_{z\in\zoon}|\psi(z)| =1, \\
&\sum_{z\in\zoon}\psi(z)f(z) > \epsilon.
\end{align*}
\label{fact:berr-morth}
\end{fact}

\noindent 
See~\cite{sherstov07quantum} for a proof of this fact using
linear programming duality.  The crux of the method
is the following theorem.

\begin{theorem}[\cite{sherstov07ac-majmaj}]
Fix a function $h:\zoon\to\moo$ and a probability distribution $\mu$
on $\zoon$ such that 
\begin{align*}
\widehat{h\circ \mu}(S)=0 && \text{~~for~~} |S|<d.
\end{align*}
Let $N$ be a given integer. Define 
\[H=[h(x|_V)]_{x,V}, \qquad P=2^{-N+n}
{N \choose n}^{-1}[\mu(x|_V)]_{x,V},\]
where the rows are indexed
by $x\in\zoo^N$ and columns by $V\in{[N]\choose n}.$ Then
\[\disc_P(H) \leq \left( \frac{4\e n^2}{N d} \right)^{d/2}.  \]
\label{thm:deg2disc}
\end{theorem}

At last, we are ready to state the pattern matrix method. 

\begin{theorem}[\cite{sherstov07quantum}]
Let $f:\zoon\to\moo$ be a given function, $d=\adeg(f).$ 
Let $N$ be a given integer. Define $F=[f(x|_V)]_{x,V},$ 
where the rows are indexed by $x\in\zoo^N$ and columns by
$V\in{[N]\choose n}.$ If $N\geq 16\e n^2/d,$ then 
\[R(F)=\Omega\left(d\log\left\{\frac{Nd}{4\e n^2}\right\}\right). \]
\label{thm:2party-pattern-matrix}
\end{theorem}

\begin{proof}[Proof \emph{(adapted from \cite{sherstov07quantum})}.]
Let $\epsilon=1/10.$ By Fact~\ref{fact:berr-morth}, there exists a function $h:\zoon\to\moo$
and a probability distribution $\mu$ on $\zoon$ such that
\begin{align}
\widehat{h\circ \mu}(S)=0,    &&|S|<d,
\label{eqn:hmu-orthog}
\end{align}
and
\begin{align}
\sum_{z\in\zoon}f(z)\mu(z)h(z) > \frac13. && \text{~~}
\label{eqn:hmu-correl}
\end{align}
Letting $H=[h(x|_V)]_{x,V}$ and $P=2^{-N+n} {N \choose
n}^{-1}[\mu(x|_V)]_{x,V},$ we obtain from~(\ref{eqn:hmu-orthog})
and Theorem~\ref{thm:deg2disc} that
\begin{align}
\disc_P(H) \leq \left( \frac{4\e n^2}{N d} \right)^{d/2}.  
\label{eqn:PH-disc}
\end{align}
At the same time, one sees from (\ref{eqn:hmu-correl}) that
\begin{align}
\langle F, H\circ P\rangle > \frac13.
\label{eqn:FHP-correl}
\end{align}
The theorem now follows from (\ref{eqn:PH-disc}) and
(\ref{eqn:FHP-correl}) in view of the generalized discrepancy method,
Theorem~\ref{thm:2party-generalized-disc}.
\end{proof}

\begin{remark*}
Presented above is a weaker, combinatorial version of the pattern
matrix method. 
The communication lower bounds in 
Theorems~\ref{thm:deg2disc} and~\ref{thm:2party-pattern-matrix}
were improved to optimal in~\cite{sherstov07quantum} 
using matrix-analytic techniques. 
Unlike the
combinatorial argument above, however, the matrix-analytic proof is not known to extend to the multiparty model and is not used in the follow-up multiparty
papers~\cite{arkadev07multiparty,lee-shraibman08disjointness,chatt-ada08disjointness,pitassi08np-rp,david-pitassi-viola08bpp-np,beame-huyn-ngoc08multiparty-circuits} or our work.

An alternate technique based on 
Fact~\ref{fact:berr-morth} is the \emph{block-composition
method}~\cite{shi-zhu07block-composed}, developed independently 
of the pattern matrix method.  
See~\cite[\S5.3]{dual-survey} for a comparative
discussion.
\end{remark*}

\section{A New Criterion for Nondeterministic and Merlin-\\Arthur Complexity
}\label{sec:newmethod}

In this section, we derive a new criterion for high
communication complexity in the nondeterministic and
Merlin-Arthur models. This criterion, inspired by the
generalized discrepancy method, will allow us to obtain
our main result.

\begin{theorem}
Let $F:X\to\moo$ be given, where $X=X_1\times\cdots\times X_k.$
Fix a function $H:X\to\moo$ and a probability distribution $P$
on $X.$ Put 
\begin{align*}
\alpha &= P(F^{-1}(-1)\cap H^{-1}(-1)), \\
\beta &= P(F^{-1}(-1)\cap H^{-1}(+1)), \\
Q &=\log \frac{\alpha}{\beta + \disc_P (H)}.
\end{align*}
Then
\begin{align}
N(F)   &\geq Q
\label{eqn:n-method}
\intertext{and}
\ma(F) &\geq 
\min\left\{\Omega(\sqrt Q), \;
\Omega\left( \frac Q{\log\{2/\alpha\}}\right)
\right\}.
\label{eqn:ma-method}
\end{align}
\label{thm:nondet-method}
\end{theorem}

\begin{proof}
Put $c=N(F).$ Then there is a cover of $F^{-1}(-1)$ by $2^c$ cylinder
intersections, each contained in $F^{-1}(-1).$ Fix one such cover,
$\chi_1,\chi_2,\dots,\chi_{2^c}:X\to\zoo.$
By the definition of discrepancy,
\begin{align*}
\left\langle \textstyle\sum \chi_i, -H\circ P\right\rangle 
\leq 2^{c}\disc_P (H).
\end{align*}
On the other hand, $\sum \chi_i$ ranges between $1$
and $2^c$ on $F^{-1}(-1)$ and vanishes on $F^{-1}(+1).$ Therefore,
\begin{align*}
\left\langle \textstyle\sum \chi_i, -H\circ P\right\rangle 
\geq \alpha - 2^{c}\beta. 
\end{align*}
These two inequalities force (\ref{eqn:n-method}).

We now turn to the Merlin-Arthur model. Let $c=\ma(F)$ and
$\delta=\alpha2^{-c-1}.$ The first step is to improve the error probability
of the Merlin-Arthur protocol by repetition from $1/3$ to
$\delta.$ Specifically, following Klauck~\cite{klauck03rect} we
observe that there exist  randomized protocols $F_1,\dots,F_{2^c}
:X\to\zoo,$ each a random variable of the coin tosses and each
having communication cost $c'=O(c\log\{1/\delta\}),$ such that the
sum
\begin{align*}
\sum \Exp[F_i]
\end{align*}
ranges in $[1-\delta,2^c]$ on $F^{-1}(-1)$ and in $[0,\delta 2^{c}]$
on $F^{-1}(+1).$ As a result,
\begin{align}
\left\langle \sum \Exp[F_i], -H\circ P\right\rangle
\geq \alpha(1-\delta) - \beta 2^c - (1-\alpha-\beta)\delta2^c.
\label{eqn:long-equation}
\end{align}
At the same time, 
\begin{align}
\left\langle \sum \Exp[F_i], -H\circ P\right\rangle
\leq \sum_{i=1}^{2^c} 2^{c'}\disc_P(H)
= 2^{c+c'}\disc_P(H).
\label{eqn:long-equation2}
\end{align}
The bounds in (\ref{eqn:long-equation}) and (\ref{eqn:long-equation2})
force (\ref{eqn:ma-method}).
\end{proof}

Since sign tensors $H$ and $-H$ have the same discrepancy under
any given distribution, we have the following
alternate form of Theorem~\ref{thm:nondet-method}.

\begin{corollary}
Let $F:X\to\moo$ be given, where $X=X_1\times\cdots\times X_k.$
Fix a function $H:X\to\moo$ and a probability distribution $P$
on $X.$ Put 
\begin{align*}
\alpha &= P(F^{-1}(+1)\cap H^{-1}(+1)), \\
\beta &= P(F^{-1}(+1)\cap H^{-1}(-1)), \\
Q &=\log \frac{\alpha}{\beta + \disc_P (H)}.
\end{align*}
Then
\begin{align*}
N(-F)   &\geq Q
\intertext{and}
\ma(-F) &\geq 
\min\left\{\Omega(\sqrt Q), \;
\Omega\left( \frac Q{\log\{2/\alpha\}}\right)
\right\}.
\end{align*}
\label{cor:nondet-method}
\end{corollary}

At first glance, it is unclear how the nondeterministic bound of
Theorem~\ref{thm:nondet-method} and its counterpart
Corollary~\ref{cor:nondet-method} relate to the generalized discrepancy
method. We now pause to make this relationship quite explicit.
Recall that nondeterminism is a kind of randomized computation,
viz., a nondeterministic protocol with cost~$c$ for a function $F$
is a kind of cost-$c$ randomized protocol with error probability
at most $\epsilon=\frac12-2^{-c}$ on $F^{-1}(-1)$ and error probability
$\epsilon=0$ elsewhere.  This is the setting of
Theorem~\ref{thm:nondet-method}.  The generalized discrepancy method,
on the other hand, has a single error parameter $\epsilon$ for all
inputs.  To best convey this distinction between the two methods,
we formulate a more general criterion yet, which allows
for different errors on each input.

\begin{theorem}
Let $F:X\to\moo$ be given, where $X=X_1\times\cdots\times X_k.$ Let
$c$ be the least cost of a public-coin protocol for $F$ with error
probability $E(x)$ on $x\in X,$ for some $E:X\to[0,1/2].$ Then for
all functions $H:X\to\moo$ and all probability distributions $P$ on
$X,$
\begin{align*}
2^c \geq \frac{\langle F,H\circ P\rangle - 2\langle P,E\rangle}
{\disc_P(H)}.
\end{align*}
\label{thm:new-multi-generalized-disc}
\end{theorem}

\begin{proof}
A public-coin protocol with cost $c$ is a probability distribution
on deterministic protocols with cost at most~$c.$ Then by hypothesis, there
are random variables
$\boldchi_1,\boldchi_2,\dots,\boldchi_{2^c}:X\to\zoo,$
each a cylinder intersection, and random variables
$\boldsigma_1,\boldsigma_2,\dots,\boldsigma_{2^c}\in\moo,$
such that 
\[ \left|F(x) - \Exp\left[\sum\boldsigma_i\boldchi_i(x)\right]
\right|\leq 2E(x) \qquad \text{~~for~~} x\in X. \]
Therefore, 
\begin{align*}
\left\langle F- \Exp\left[\sum\boldsigma_i\boldchi_i\right], H\circ P
\right\rangle
&\leq 2\langle P, E\rangle.
\end{align*}
On the other hand,
\begin{align*}
\left\langle F- \Exp\left[\sum\boldsigma_i\boldchi_i\right],
H\circ P \right\rangle
&\geq \langle F,H\circ P\rangle - 2^{c}\disc_P(H)
\end{align*} 
by the definition of discrepancy.
The theorem follows at once from the last two inequalities.
\end{proof}

\section{Main Result}  \label{sec:main}

We now prove the claimed separations of
nondeterministic, co-nondeterministic, and
Merlin-Arthur communication complexity.  It will be
easier to first obtain these separations by a probabilistic
argument and only then sketch an explicit construction.

We start by deriving a suitable analytic property of the {\sc or}
function.

\begin{theorem}
There is a function $\psi:\zoom\to\Re$ such that:
\begin{align}
\sum_{z\in\zoom}|\psi(z)| &=1,
\label{eqn:new-psi-bounded} \\
\hat\psi(S)&=0 &&\text{for~} |S|\leq\Theta(\sqrt m),
\label{eqn:new-psi-orthog}\\
\psi(0)&> \frac16.
\label{eqn:new-psi-correl}
\end{align}
\label{thm:or-property}
\end{theorem}

\begin{proof}
Let $f:\zoom\to\moo$ be given by $f(z)=1\Leftrightarrow z=0.$
It is well-known~\cite{nisan-szegedy94degree,paturi92approx} that
$\adeg(f)\geq\Omega(\sqrt m).$ By Fact~\ref{fact:berr-morth}, there is
a function $\psi:\zoom\to\Re$ that
obeys~(\ref{eqn:new-psi-bounded}),~(\ref{eqn:new-psi-orthog}), and
additionally satisfies
\begin{align*}
\sum_{z\in\zoom}\psi(z)f(z)>\frac13.
\end{align*}
Finally,
\begin{align*}
2\psi(0) = \sum_{z\in\zoom} \psi(z)\{f(z)+1\}
= \sum_{z\in\zoom} \psi(z)f(z)
> \frac13,
\end{align*}
where the second equality follows from $\hat\psi(\emptyset)=0$. 
\end{proof}

For the remainder of this section, it will be convenient to
establish some additional notation following David and
Pitassi~\cite{pitassi08np-rp}.  Fix integers $n,m$ with
$n>m.$ Let $\psi:\zoom\to\Re$ be a given function with
$\sum_{z\in\zoom}|\psi(z)|=1.$ Let $d$ denote the least order of a
nonzero Fourier coefficient of $\psi.$ Fix a Boolean function
$h:\zoo^m\to\moo$ and the distribution $\mu$ on $\zoom$ such that
$\psi(z)\equiv h(z)\mu(z).$ For a mapping $\alpha:(\zoon)^k\to{[n]\choose
m},$ define a $(k+1)$-party communication problem
$H_\alpha:(\zoon)^{k+1}\to\moo$ by $H_\alpha(x,y_1,\dots,y_k) =
h(x|_{\alpha(y_1,\dots,y_k)}).$ 
Define a distribution $P_\alpha$ on $(\zoon)^{k+1}$
by $P_\alpha(x,y_1,\dots,y_k) =
2^{-(k+1)n+m}\mu(x|_{\alpha(y_1,\dots,y_k)}).$
The following theorem combines the pattern matrix
method with a probabilistic argument.

\begin{theorem}[\cite{pitassi08np-rp}]
Assume that $n\geq 16\e m^2 2^k.$ Then for a uniformly random choice
of $\alpha:(\zoon)^k\to{[n] \choose m},$
\[ \Exp_\alpha \left[\disc_{P_\alpha}(H_\alpha)^{2^k} \right]
\leq 2^{-n/2} + 2^{-d2^k+1}.\]
\label{thm:pitassi}
\end{theorem}

For completeness, we include a detailed proof of this result.
\begin{proof}[Proof \emph{(reproduced from the survey
article~\cite{dual-survey}, pp.~88--89).}]
By Theorem~\ref{thm:disc-bound},
\begin{align}
\disc_{P_\alpha}(H_\alpha)^{2^k} \leq 2^{m2^k}
\Exp_{Y} |\Gamma(Y)|,  \label{eqn:disc-bound-pitassi}
\end{align}
where we put
$Y=(y_1^0,y_1^1,\dots,y_{k}^0,y_{k}^1)\in(\zoon)^{2k}$
and 
\[ \Gamma(Y)=
\Exp_{x}\left[
\prod_{z\in\zook}
\psi\left(x|_{\alpha\left(y_1^{z_1},y_2^{z_2},\dots,y_k^{z_k}\right)}\right)
\right]. \]
For a fixed choice of $\alpha$ and $Y$, we will use the shorthand $S_z=\alpha(y_1^{z_1},\dots,y_k^{z_k}).$ 
To analyze $\Gamma(Y),$ one proves two key claims analogous
to those in the two-party Theorem~\ref{thm:deg2disc}
(see~\cite{sherstov07ac-majmaj,dual-survey} for more detail). 
\begin{claim}
Assume that $\left|\bigcup_{z\in\zook} S_z\right|> m2^k-d2^{k-1}.$ Then $\Gamma(Y) = 0.$ 
\label{cla:low-terms-pitassi}
\end{claim}

\begin{proof}
If $\left|\bigcup S_z\right|>m2^k-d2^{k-1},$ then some $S_z$ must feature more
than $m-d$ elements that do not occur in $\bigcup_{u\ne z} S_u.$
But this forces $\Gamma(Y)=0$ since the Fourier transform of $\psi$
is supported on characters of order $d$ and higher.
\end{proof}

\begin{claim}
For every $Y$, $|\Gamma(Y)| \leq 2^{-\left|\cup S_z\right|}$.
\label{cla:high-terms-pitassi}
\end{claim}

\begin{proof}
Immediate from Proposition~\ref{prop:distribution}.
\end{proof}

In view of (\ref{eqn:disc-bound-pitassi}) and
Claims~\ref{cla:low-terms-pitassi} and~\ref{cla:high-terms-pitassi},
we have 
\[ \Exp_{\alpha} \left[\disc_{P_{\alpha}}(H_\alpha)^{2^k}\right]
\leq \sum_{i=d2^{k-1}}^{m2^k-m} 2^{i} 
\Prob_{Y,\alpha}\left[\left|\bigcup S_z\right|=m2^k-i\right]. \]
It remains to bound the 
probabilities in the last expression. 
With probability at least $1-k2^{-n}$ over the choice of $Y$,
we have $y_i^0\ne y_i^1$ for each $i=1,2,\dots,k.$
Conditioning on this event, the fact that $\alpha$
is chosen uniformly at random means that the $2^k$ sets $S_z$
are distributed independently and uniformly over
${[n]\choose m}.$ A calculation now reveals that 
\[\Prob_{Y,\alpha}\left[\left|\bigcup S_z\right|=m2^k-i\right]\leq
k2^{-n} +  {m2^k\choose i} \left(\frac{m2^k}{n}\right)^i
\leq k2^{-n} +  8^{-i}.
\qedhere \]
\end{proof}

We are ready to prove our main result.  It may be helpful to
contrast the proof to follow with the proof of the pattern matrix
method (Theorem~\ref{thm:2party-pattern-matrix}).

\begin{theorem}
Let $k\leq (1-\epsilon)\log n,$ where $\epsilon>0$ is any 
given constant.  Then there exists a function
$F_\alpha:(\zoon)^{k+1}\to\moo$ such that:
\begin{align}
N(F_\alpha)&=O(\log n)        \label{eqn:nondet-low}
\end{align}
and
\begin{align}
\ma(-F_\alpha)&= n^{\Omega(1)}. \label{eqn:conondet-high}
\end{align}
In particular, $\coNP_k^{cc}\nsubseteq\MA_k^{cc}$ and $\NP_k^{cc}\ne\coNP_k^{cc}.$
\label{thm:np-conp-existential}
\end{theorem}

\newcommand{\OR}{\text{\textsc{or}}}

\begin{proof}
Let $m=\lfloor n^{\delta}\rfloor$ for a sufficiently small constant
$\delta=\delta(\epsilon)>0.$ As usual, define
$\OR_m:\zoom\to\moo$ by $\OR_m(z)=1\Leftrightarrow z=0.$ 
Let $\psi:\zoo^m\to\Re$ be as guaranteed by Theorem~\ref{thm:or-property}.
For a mapping
$\alpha:(\zoon)^k\to{[n]\choose m},$ let $H_\alpha$ and $P_\alpha$
be defined in terms of $\psi$ as described earlier in this section.
Then Theorem~\ref{thm:pitassi} shows the existence of $\alpha$ such
that
\begin{align}
\disc_{P_\alpha}(H_\alpha) \leq 2^{-\Omega(\sqrt m)}.
\label{eqn:discpalpha}
\end{align} 
Define $F_\alpha:(\zoon)^{k+1}\to\moo$ by
$F_\alpha(x,y_1,\dots,y_k) = \OR_m(x|_{\alpha(y_1,\dots,y_k)}).$
It is immediate from the properties of $\psi$ that
\begin{align}
P_\alpha(F_\alpha^{-1}(+1)\cap H_\alpha^{-1}(+1)) > \frac16,
\label{eqn:minusones}\\
\rule{0mm}{6mm}
P_\alpha(F_\alpha^{-1}(+1)\cap H_\alpha^{-1}(-1)) = 0.
\label{eqn:plusones}
\end{align}
The sought lower bound in (\ref{eqn:conondet-high}) now follows from
(\ref{eqn:discpalpha})--(\ref{eqn:plusones}) and
Corollary~\ref{cor:nondet-method}.

On the other hand, as observed in~\cite{pitassi08np-rp},
the function $F_\alpha$ has an efficient nondeterministic protocol.
Namely, player~$1$ (who knows $y_1,\dots,y_k$) nondeterministically
selects an element $i\in \alpha(y_1,\dots,y_k)$ and writes $i$ on
the shared blackboard.  Player~$2$ (who knows $x$) then announces
$x_i$ as the output of the protocol. This yields the desired upper
bound in~(\ref{eqn:nondet-low}).
\end{proof}

As promised, we will now sketch an explicit construction of the
function whose existence has just been proven. For this, it suffices to
invoke previous work by David, Pitassi, and
Viola~\cite{david-pitassi-viola08bpp-np}, who derandomized the
choice of $\alpha$ in Theorem~\ref{thm:pitassi}. More precisely,
instead of working with a family $\{H_\alpha\}$ of functions, each
given by $H_\alpha(x,y_1,\dots,y_k) = h(x|_{\alpha(y_1,\dots,y_k)}),$
the authors of~\cite{david-pitassi-viola08bpp-np} posited a single
function $H(\alpha,x,y_1,\dots,y_k)= h(x|_{\alpha(y_1,\dots,y_k)}),$
where the new argument $\alpha$ is known to all players and ranges
over a small, explicitly given subset $A$ of all mappings
$(\zoon)^k\to{[n] \choose m}.$ By choosing $A$ to be
pseudorandom, the authors of~\cite{david-pitassi-viola08bpp-np}
forced the same qualitative conclusion in Theorem~\ref{thm:pitassi}. This
development carries over unchanged to our setting, and we obtain
our main result.

\begin{restatetheorem}{thm:main}
\mainthe
\end{restatetheorem}

\begin{proof}
Identical to Theorem~\ref{thm:np-conp-existential}, with the
described derandomization of $\alpha.$
\end{proof}

\section{On Disjointness and Constant-Depth Circuits}

In this final section, we revisit recent multiparty analyses of 
the disjointness function and other constant-depth
circuits~\cite{arkadev07multiparty,lee-shraibman08disjointness,chatt-ada08disjointness,beame-huyn-ngoc08multiparty-circuits}.
We will see that the program of the previous sections applies essentially
unchanged to these other functions.

We start with some notation.  Fix a function $\phi:\zoom\to\Re$ and
an integer $N$ with $m\mid N.$ Define
the \emph{$(k,N,m,\phi)$-pattern tensor} as the $k$-argument function
$A:\zoo^{m(N/m)^{k-1}}\times[N/m]^m\times\cdots\times [N/m]^m\to\Re$ given by
$A(x,V_1,\dots,V_{k-1})= \phi(x|_{V_1,\dots,V_{k-1}}),$ where 
\[ 
x|_{V_1,\dots,V_{k-1}}=
\left(x_{1,V_1[1],\dots,V_{k-1}[1]},
\; \dots,\;
x_{m,V_1[m],\dots,V_{k-1}[m]}\right)\in\zoom
\]
and $V_j[i]$ denotes the $i$th element of the $m$-dimensional vector $V_j.$
(Note that we index the string $x$ by viewing it as a $k$-dimensional array
of $m\times(N/m)\times\cdots\times(N/m)=m(N/m)^{k-1}$ bits.)
This definition extends \emph{pattern
matrices}~\cite{sherstov07ac-majmaj,sherstov07quantum} to higher dimensions. 
The two-party Theorem~\ref{thm:deg2disc} 
has been adapted as follows to $k\geq3$ players.

\begin{theorem}
[\cite{arkadev07multiparty,lee-shraibman08disjointness,chatt-ada08disjointness}]
Fix a function $h:\zoom\to\moo$ and a probability distribution $\mu$
on $\zoom$ such that \[\widehat{h\circ \mu}(S)=0,\qquad |S|<d.\]
Let $N$ be a given integer, $m\mid N.$ Let $H$ be the $(k,N,m,h)$-pattern
tensor. Let $P$ be the $(k,N,m,2^{-m(N/m)^{k-1}+m}
(N/m)^{-m(k-1)}\mu)$-tensor.
If $N\geq 4\e m^2(k-1)2^{2^{k-1}}/d,$ then 
\[\disc_P(F) \leq 2^{-d/2^{k-1}}.\]
\label{thm:multiparty-deg2disc}
\end{theorem}

\noindent
A proof of this exact formulation is available in the survey
article~\cite{dual-survey}, pp.~85--86.  We are now prepared to
apply our techniques to the disjointness function.

\begin{theorem}
Let $N$ be a given integer, $m\mid N.$ Let $F$ be the $(k,N,m,\OR_m)$-pattern
tensor.  
If $N\geq 4\e m^2(k-1)2^{2^{k-1}}/d,$ then 
\begin{align*}
N(-F) \geq \Omega\left(\frac{\sqrt m}{2^k}\right),
\qquad
\ma(-F) \geq \Omega\left(\frac{\sqrt[4] m}{2^{k/2}}\right).
\end{align*}
\label{thm:pattern-tensor}
\end{theorem}

\begin{proof}
Let $\psi:\zoo^m\to\Re$ be as guaranteed by Theorem~\ref{thm:or-property}.
Fix a function $h:\zoom\to\moo$ and a distribution $\mu$ on $\zoom$
such that $\psi(z)\equiv h(z)\mu(z).$ Let $H$ be the
$(k,N,m,h)$-pattern tensor. Let $P$ be the $(k,N,m,2^{-m(N/m)^{k-1}+m}
(N/m)^{-m(k-1)}\mu)$-pattern tensor, which is a probability
distribution.  Then by Theorem~\ref{thm:multiparty-deg2disc},
\begin{align}
\disc_{P}(H) \leq 2^{-\Omega(\sqrt m/2^k)}.
\label{eqn:disctensor}
\end{align} 
On the other hand, it is clear from the properties of $\psi$ that
\begin{align}
P(F^{-1}(+1)\cap H^{-1}(+1)) > \frac16,
\label{eqn:minusones-tensor}\\
\rule{0mm}{6mm}
P(F^{-1}(+1)\cap H^{-1}(-1)) = 0.
\label{eqn:plusones-tensor}
\end{align}
In view of 
(\ref{eqn:disctensor})--(\ref{eqn:plusones-tensor}) and
Corollary~\ref{cor:nondet-method},
the proof is complete.
\end{proof}

The function $F$ in Theorem~\ref{thm:pattern-tensor} is a
subfunction of the multiparty disjointness function $\text{\sc
disj}:(\zoon)^k\to\moo,$ where $n=m(N/m)^{k-1}$ and 
\[ \text{\sc disj}(x_1,\dots,x_k) = \bigvee_{j=1}^n \bigwedge_{i=1}^k
x_{ij}. \]
Recall that disjointness has trivial nondeterministic complexity,
$O(\log n).$ In particular, Theorem~\ref{thm:pattern-tensor} shows
that the disjointness function separates $\NP_{k}^{cc}$ from $\coNP_k^{cc}$ and witnesses that $\coNP_k^{cc}\nsubseteq\MA_k^{cc}$ for up to $k=\Theta(\log\log n)$ players.  
Our technique similarly applies to the follow-up work on disjointness by Beame and Huynh-Ngoc~\cite{beame-huyn-ngoc08multiparty-circuits}, whence we obtain the stronger consequence that the disjointness function separates $\NP_k^{cc}$ from $\coNP_k^{cc}$ and witnesses that $\coNP_k^{cc}\nsubseteq\MA_k^{cc}$ for up to $k=\Theta(\log^{1/3} n)$ players.

We conclude this section with a remark on constant-depth circuits.
Let $\epsilon$ be a sufficiently small absolute constant, 
$0<\epsilon<1.$ For each $k=2,3,\dots,\epsilon\log n,$
the authors of~\cite{beame-huyn-ngoc08multiparty-circuits} 
construct a constant-depth circuit $F:(\zoon)^k\to\moo$ with
$N(F)=\log^{O(1)} n$ and $R(F)=n^{\Omega(1)}.$ A glance
at the proof in~\cite{beame-huyn-ngoc08multiparty-circuits} reveals,
once again, that the program of our paper is readily applicable to $F,$
with the consequence that \mbox{$\ma(-F)=n^{\Omega(1)}$}.
In particular, our work shows that $\NP^{cc}_k \neq \coNP^{cc}_k$ and $\coNP_k^{cc}\nsubseteq\MA_k^{cc}$ for up to $k=\epsilon\log n$ players, as witnessed by a constant-depth circuit.

{\small
\bibliographystyle{alphanumm}
\bibliography{bib/general,bib/fourier,bib/cc,bib/learn,bib/sherstov}
}

\end{document}